%% file: Icrc_proceedings_2025.tex
\documentclass[a4paper,11pt]{article}

\usepackage{pos}
\usepackage{subcaption}
\usepackage{graphicx}
\usepackage{rotating}
\usepackage{wrapfig}
\usepackage{gensymb}
\usepackage{lineno}
\usepackage[inkscapelatex=false]{svg}
 
\newcommand{\refer}[1]{Ref.~\cite{#1}}
\usepackage{lipsum}

\title{Radio detection of Cosmic Rays with the IceCube Surface Array Enhancement}

\ShortTitle{Radio detection of Cosmic Rays with the IceCube Surface Array Enhancement}

\author{The IceCube Collaboration \\{\normalsize \normalfont(a complete list of authors can be found at the end of the proceedings)}\\}

\emailAdd{megha.venugopal@kit.edu}

\abstract{

The surface array of the IceCube Neutrino Observatory, IceTop, measures cosmic rays in the PeV-EeV primary energy range. Stations comprising radio antennas and scintillation detectors will be added to enhance the existing surface detectors. A prototype station, consisting of eight scintillation detectors and three radio antennas, has been in operation in with the instrumentation in final design since the beginning of 2023. Radio signals from air showers are measured by antennas that are read-out when the trigger condition from the scintillation detectors is met. This contribution reports on air-shower coincidence measurements of these radio antennas with IceTop. Geometric shower parameters reconstructed from the radio antennas are compared with those from IceTop to determine the angular resolution. We also present details on the two new stations that were tested, deployed and commissioned with their respective data acquisition systems during the latest field season at the South Pole.

\vspace{4mm}

{\bfseries Corresponding authors:}
Megha Venugopal$^{1*}$\\
{$^{1}$ \itshape Karlsruhe Institute of Technology, Institute for Astroparticle Physics}\\
[4mm]$^*$ Presenter
}
\input{ICRCdetails.tex}

\begin{document}

\maketitle

\section{Introduction}\label{sec 1}

The IceCube Neutrino Observatory is a large-scale neutrino detector built to find astrophysical neutrinos from high-energy sources. It consists of more than 5000 Digital Optical Modules (DOMs) housed in 86 strings distributed over a cubic kilometer of in ice volume. Further, seven new strings are expected to be deployed with a denser spacing to detect lower energy neutrinos as part of the IceCube Upgrade in the upcoming season.
The surface detector, IceTop, was built as a cosmic-ray detector that can also veto atmospheric neutrinos. It measures cosmic-ray-induced air showers in the PeV-EeV energy range. IceTop consists of 81 pairs of tanks filled with clear ice that detect Cherenkov light from charged particles that pass through the tanks. Due to unequal snow accumulation driven by winds on top of the tanks, the signal attenuation increases and changes every year leading to uncertainties of IceTop measurements. Progress on measures to accurately account for the effects of snow accumulation can be found in Ref.~\cite{Snowmodel,snowmodel25}. To mitigate these uncertainties and to increase threshold of air-shower measurements, 32 stations, each with 3 antennas and 8 scintillators mounted along the IceTop footprint were proposed to be built. The scintillators with a coincidence condition serve as a stand-alone detector but also act as a trigger for the radio antennas. In 2020, a prototype station with the configuration shown in Fig.~\ref{fig:Newlydeployedstations} was deployed with preliminary results presented in Ref.~\cite{HrvojeShowerIdentification} and the first results of the reconstruction of the depth of the shower maximum described in Ref.~\cite{Xmaxpaper}. The station has gone through several upgrades over the years with a change in the data acquisition to improve radio data in 2022. The scintillation detectors were upgraded to improve detector performance in 2023. The firmware for scintillation detectors was further updated to increase runtime in July 2022. Further details can be found in Ref.~\cite{ShefaliIcrc2025}.
A study of the air shower identification by calculating the signal-to-noise ratio (SNR) with the 2022 dataset is presented in Ref.~\cite{ARENAMeg}.
This contribution focuses on data from the radio antennas at the South Pole in the whole year of 2023 and the processing involved. Certain periods were excluded when the mode of data taking was different, when the data quality was affected by the presence of a windstorm and when the corresponding IceTop data was not yet processed. 
Furthermore, two new complete stations were deployed in the field season 2024-25 following the successful measurements from the prototype station. Details on the deployment are also presented in the contribution.

\section{Data Processing and air shower identification}

Radio data triggered by the coincident scintillation detectors is processed from a binary format and includes the waveform in ADC counts, the timestamps from a WhiteRabbit node (a sub nanosecond level timing synchronization system) and Region of Interest (ROI) which provides the information on the starting position of the buffer for each channel. Each antenna has two polarizations which are treated as independent channels. Each channel contains four copies of the same waveform, the median of the four traces is used for analysis. 

\begin{figure}[ht]
    \centering
	\includegraphics[width=1\textwidth]{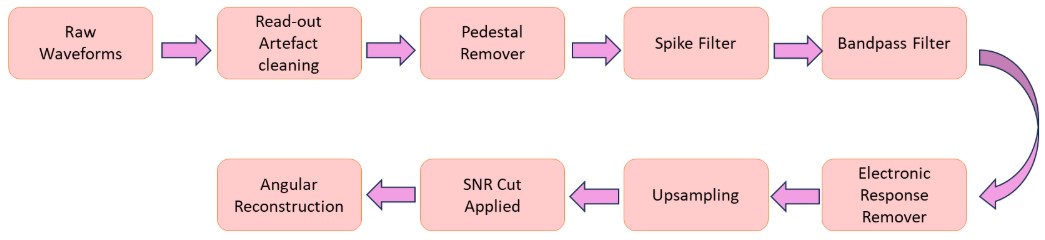}
	\protect\caption{The processing pipeline for identification of air-showers with radio antennas with a Signal-to-Noise-Ratio method (SNR).}
    \label{fig:Processingflowchart}
\end{figure}
The data is converted into the i3-file format of the IceCube framework and processed to combine data from the scintillation detectors using a coincidence window. The data from 2023 needed an additional offset correction of 1 second as a result of a firmware upgrade of the White Rabbit PPS (Pulse Per Second) before attempting a coincidence search with IceTop data.
When data from any one or both the surface stations are found to have coincidences with IceTop, then the data is combined and stored. More information on air-shower search by scintillation detectors in the year 2023 can be found in \refer{ShefaliIcrc2025}.

\begin{figure}[ht]
\begin{subfigure}{0.53\textwidth}
\centering
\includegraphics[width=1\linewidth]{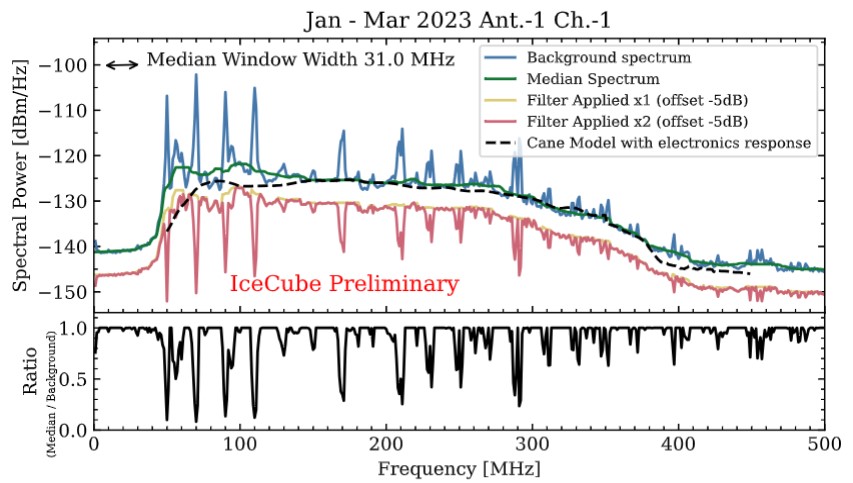}
\label{Spike}
\end{subfigure}
\hfill
\begin{subfigure}{0.45\textwidth}
\centering
\includegraphics[width=1\linewidth]{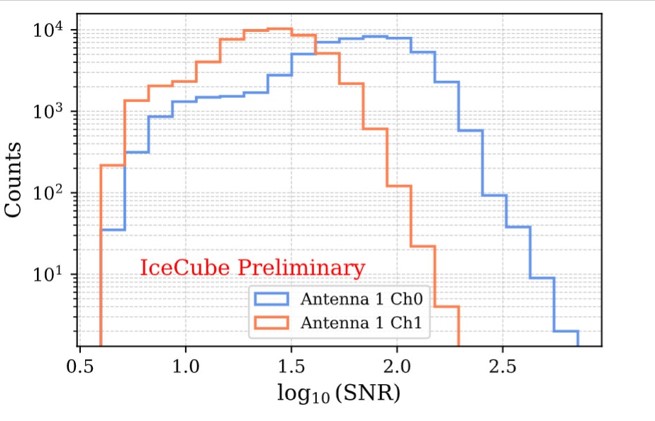}
\label{fig:SNR_bkg}
\end{subfigure}

\caption{Left: The background spectra averaged over several measured waveforms for the time period from January to March for one antenna channel. The median spectra of the frequency spectrum is also shown with a high likeness to the Cane Model of Galactic Noise folded with the electronic response. The Spike Filter is applied twice to notch those frequencies that are more likely to be noise. The ratio of the median spectra to the measured background spectra shows the exact effect of the Spike Filter. 
Right: The SNR values are computed for background data for each of the two polarizations for one antenna. 95\% of the SNR value is used as the background rejection criteria to measure air showers. The background distributions of noise varies for each antenna channel substantially since they measure different noise and are also electronically independent from each other.}
\label{fig:Spikeandbkg}
\end{figure}

The combined data is then processed through the pipeline shown in Fig.~\ref{fig:Processingflowchart} to clean the sample. A simple description can be found in \refer{ARENAMeg}. The geometry for the processing was updated to account for the extension of cables in two antennas and the changed position of one antenna in the field season 2022/23. The first analysis step is to characterize the background noise of the radio data. For this purpose, data at a fixed rate using a software trigger from the FPGA (Field-Programmable Gate Array) was used for the background. A "Spike Filter" is applied to remove noise at particular frequencies or "spikes" in the background spectrum. The ratio of the median spectrum to the measured background is used as a filter weighting those frequencies with higher noise. The filter is applied twice to notch those frequencies which are expected to be dominated by noise. The weighting method is described in \refer{ColemanFreqFilter}. For the 2023 dataset, three different spike filters were made to account for the changing background at the South Pole. The filter used for the period from January to March is shown in Fig.~\ref{fig:Spikeandbkg} (left).
The Spike Filter is applied to the data both to determine the SNR values and to identify air showers. The data is bandpassed in the 100 -- 250\,MHz frequency range based on the stability of the median background spectra. The measured electronic response is removed and the data is upsampled to reduce the probability of systematically underestimating the signal amplitude. The SNR values are computed for each month of data in 2023 and treated individually for each antenna channel. The SNR distribution of background data for one month of data is shown in Fig.~\ref{fig:Spikeandbkg}. It can be observed that the background distribution varies for each polarization and antenna. To identify air showers, events are considered if each antenna had at least one channel with a signal above the 0.95 SNR threshold value. The peak of the Hilbert envelope of the cleaned waveform after it has passed through the processing pipeline is used to determine the timing of this signal. The timing weighted with the signal amplitude is used to reconstruct the arrival direction of the shower. 

\begin{figure}[ht]
    \centering
	\includegraphics[width=1\textwidth]{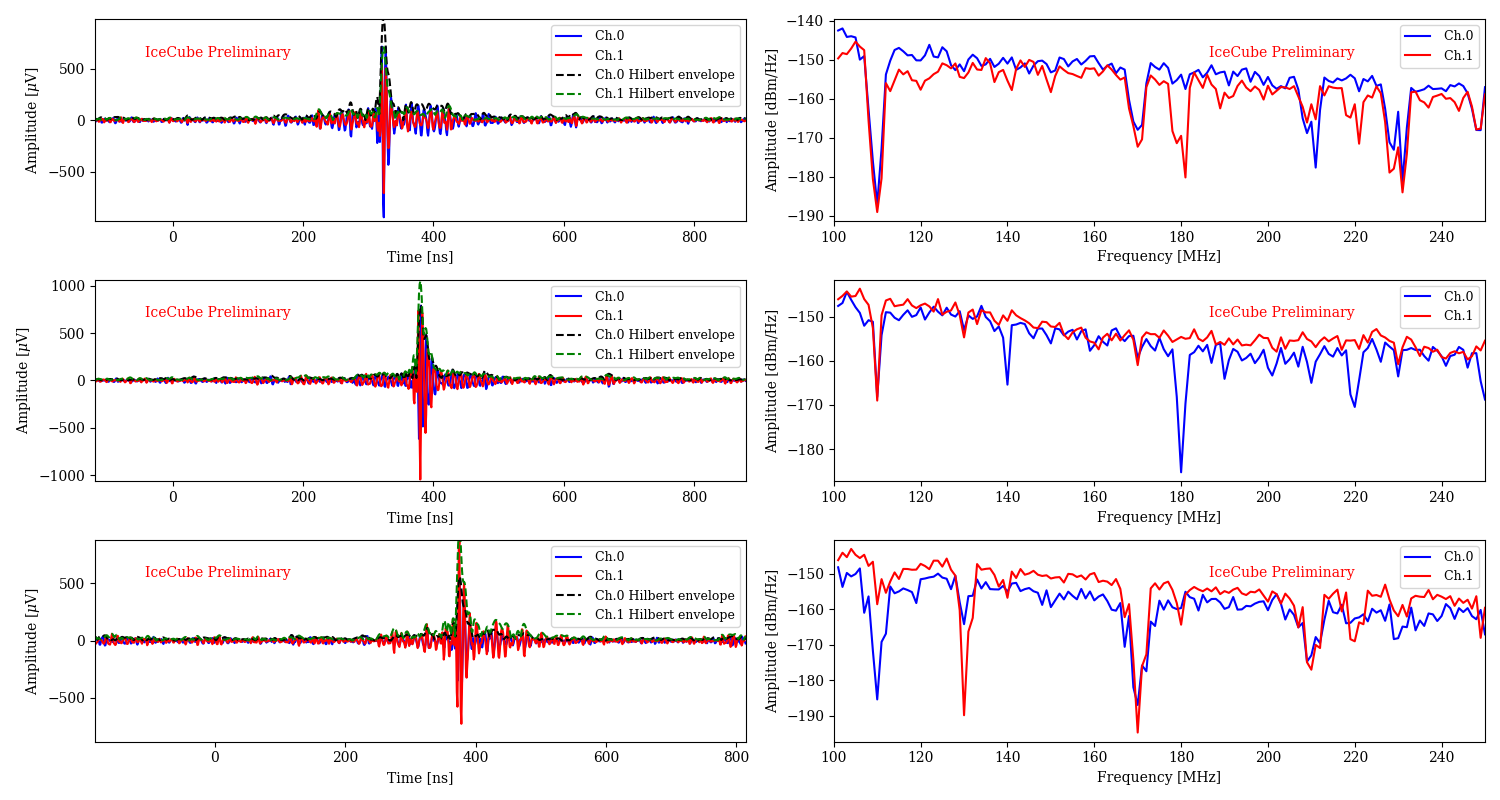}
	\protect\caption{Example of an identified radio pulse of an air shower with the Hilbert envelope overlaid (dashed line) and the  frequency spectra on the right. The dips in the frequency spectra appear as a result of the frequency notches produced by the Spike Filter.}
    \label{fig:ExampleRadio}
\end{figure}

\begin{figure}[h]
    \centering
    \begin{subfigure}{0.50\linewidth}
        \centering
        \includegraphics[width=\linewidth]{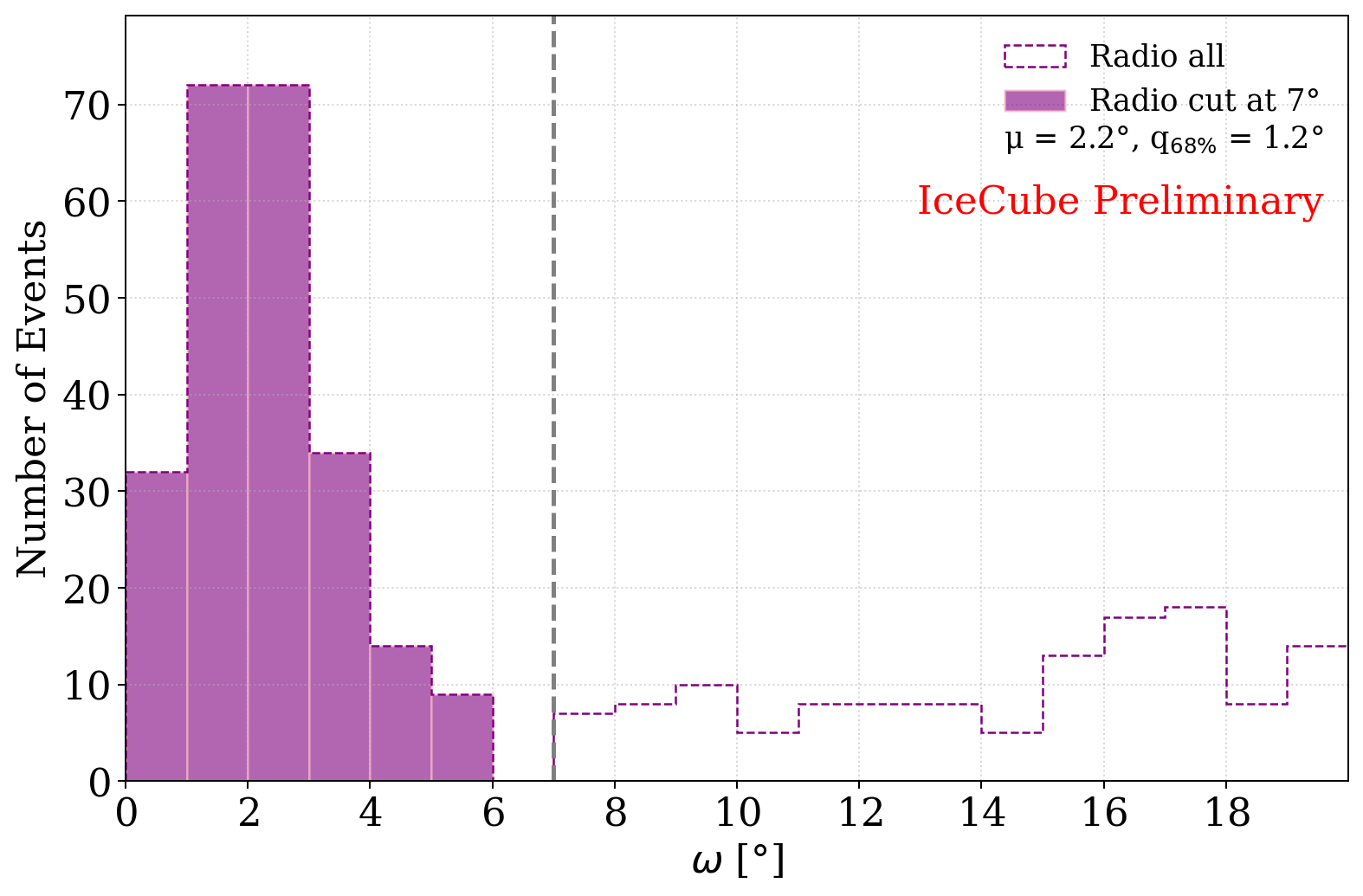}
        \caption{Opening angle distribution representing the difference in directional reconstruction of Radio antennas and IceTop}
        \label{fig:delomeg}
    \end{subfigure}
    \hfill
    \begin{subfigure}{0.48\linewidth}
        \centering
        \includegraphics[width=\linewidth]{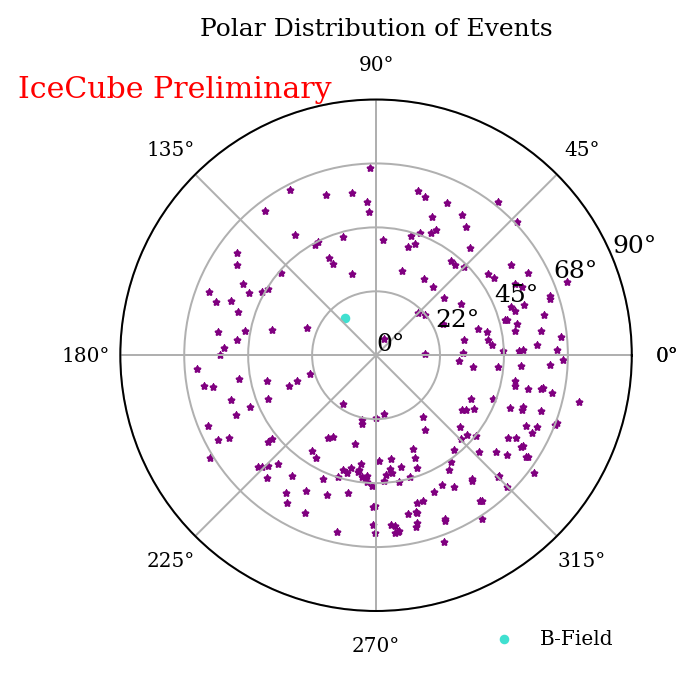}
        \caption{The distribution of events in the sky}
        \label{fig:Skydist}
    \end{subfigure}
    \caption{Left: The difference in the directional reconstruction of radio when compared to IceTop. The cut at 7 \degree\,is to select air-shower events in good agreement with IceTop. 
    Right: The sky distribution of events with an $\omega$ < 7\degree\,is presented and the blue dot indicates the B-Field at the South Pole. We observe fewer events around the B-Field accounting for the scaling of radio events with the sine of the geomagnetic angle.}
    \label{fig:DelomegSkydist}. 
\end{figure}

\begin{figure}[h]
    \centering
    \begin{subfigure}{0.50\linewidth}
        \centering
        \includegraphics[width=\linewidth]{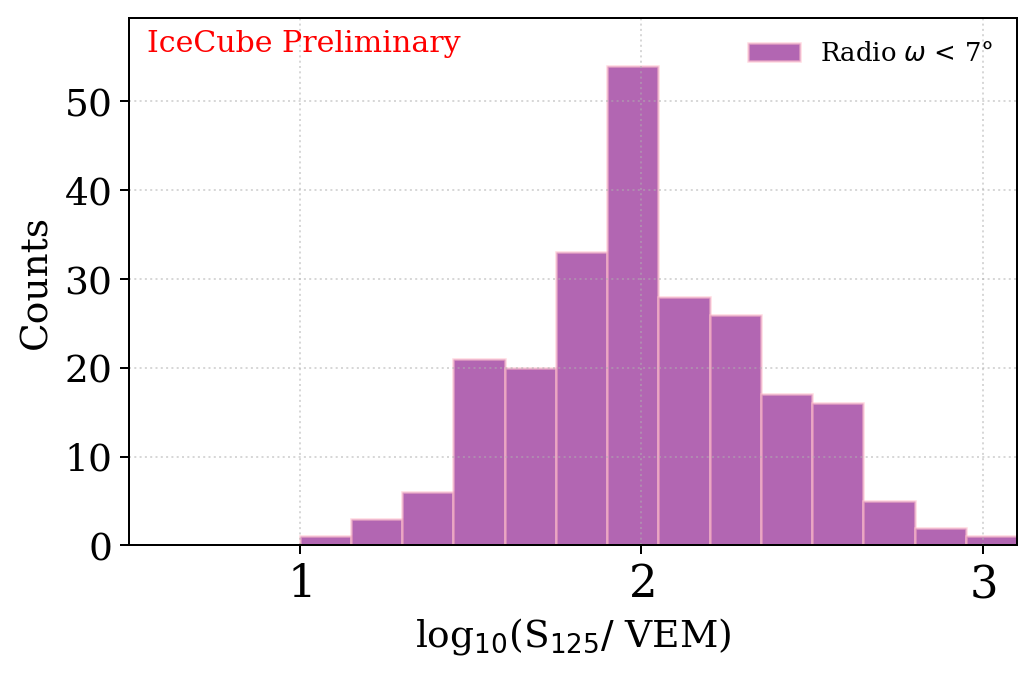}
        \caption{$S_\mathrm{125}$ distribution of all events after cut on $\omega$}
        \label{fig:S125}
    \end{subfigure}
    \hfill
    \begin{subfigure}{0.48\linewidth}
        \centering
        \includegraphics[width=\linewidth]{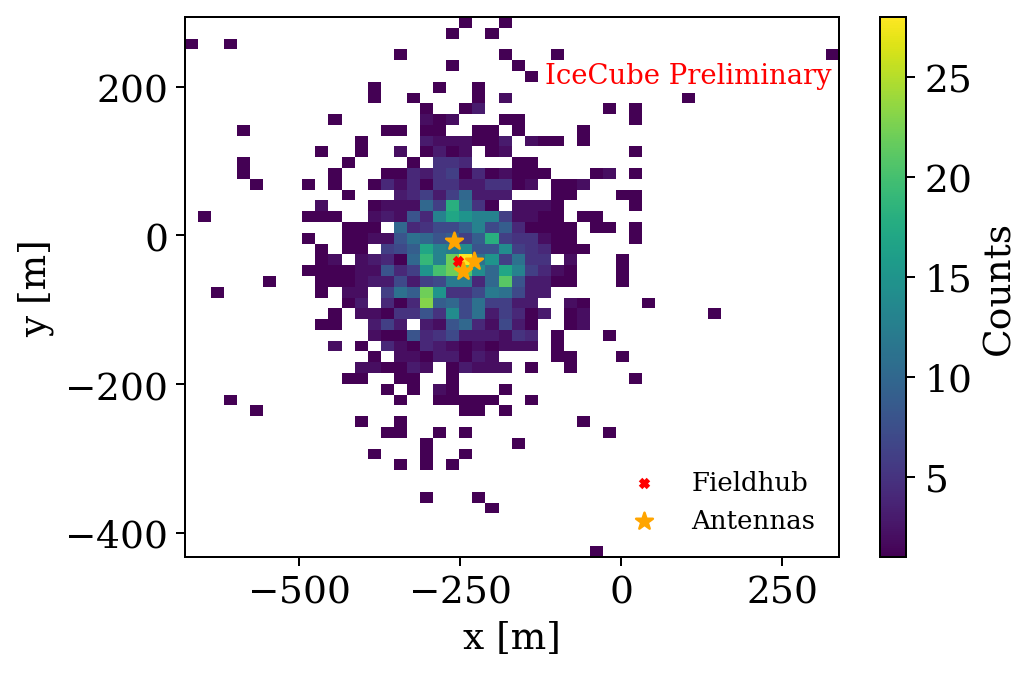}
        \caption{Core distribution from IceTop of all identified radio events}
        \label{fig:coredist}
    \end{subfigure}
    \caption{Left: The distribution of $S_\mathrm{125}$ as a primary energy proxy of the identified air-shower events after a cut on the opening angle with respect to IceTop. Right: The distribution of the core of the air showers which is a quantity derived from standard IceTop reconstruction for those events detected in radio.}
    \label{fig:S125Core}. 
\end{figure}

An example waveform of an identified radio event and its frequency spectra are given in Fig.~\ref{fig:ExampleRadio}.
 Without cuts or cleaning and after processing we identify around 1000 events. Many events were wrongly reconstructed due to the presence of an unidentified pulsed noise source in the same order of magnitude as a radio pulse within the analyzed frequency range in the trace.
 These also contribute to a reduced dataset of 233 radio events that survive the cut of having an opening angle of 7\degree\, where the opening angle is defined as the angle between the directional reconstruction of the arrival direction from radio antennas and from IceTop. The distribution of the opening angle, $\omega$, is shown in Fig.~\ref{fig:delomeg}, where events with $\omega$  > 7\degree\, are cut from the sample. 
 We use the 68\textsuperscript{th} percentile of the distribution of reconstruction differences between the two detectors as an estimate of the resolution, analogous to the 1$\sigma$ width for a Gaussian distribution. This gives us a resolution of 1.2\degree\, with respect to IceTop. The distribution of events after the cut is shown in zenith and azimuth in Fig.~\ref{fig:Skydist}. 
The quantity $S_\mathrm{125}$ serves as the energy proxy for IceTop. It is defined as the signal strength at a reference distance of 125\,m from the shower axis \cite{S125IceTop}. The $S_\mathrm{125}$ distribution of the identified events is given in Fig.~\ref{fig:S125}.
\section{2025 Field Deployment of Surface Stations}
\begin{figure}[ht]
\vspace{-0.9cm}
    \centering

    \begin{subfigure}[t]{0.55\linewidth}
        \centering
        \includegraphics[width=\linewidth]{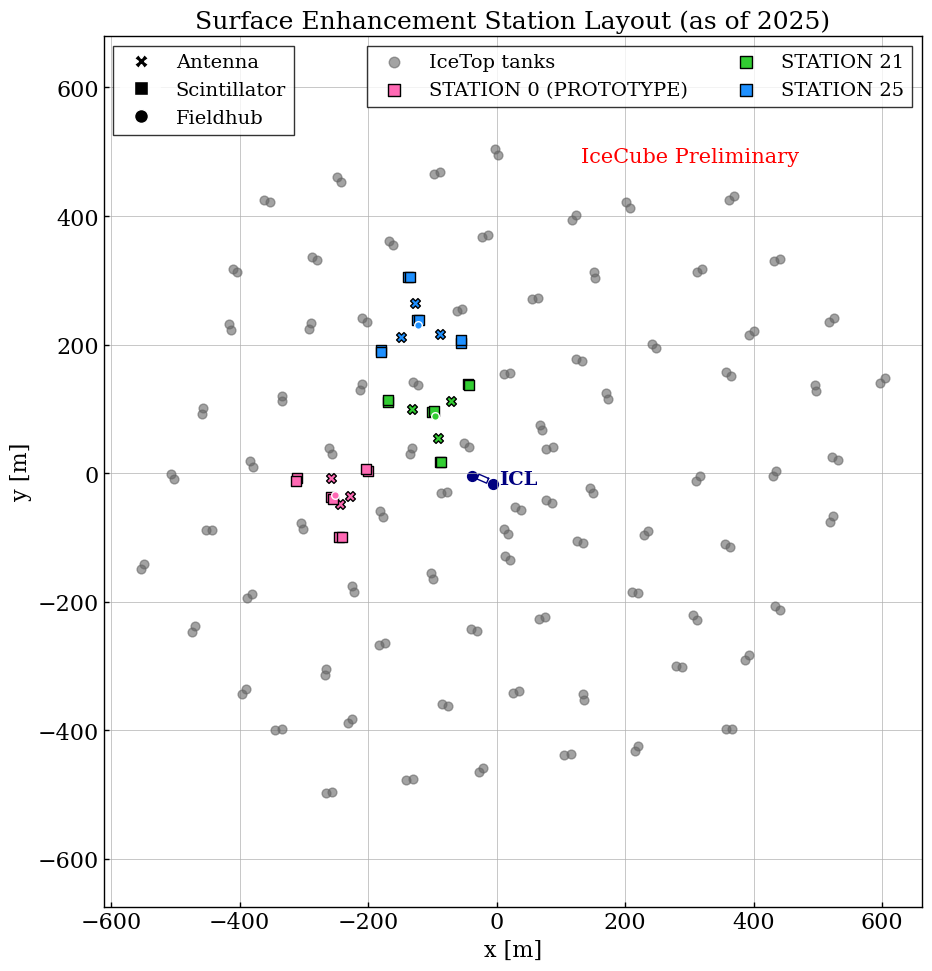}
        \label{fig:Layout}
    \end{subfigure}
    \hfill
    \begin{subfigure}{0.4\linewidth}
        \centering
        \includegraphics[width=\linewidth]{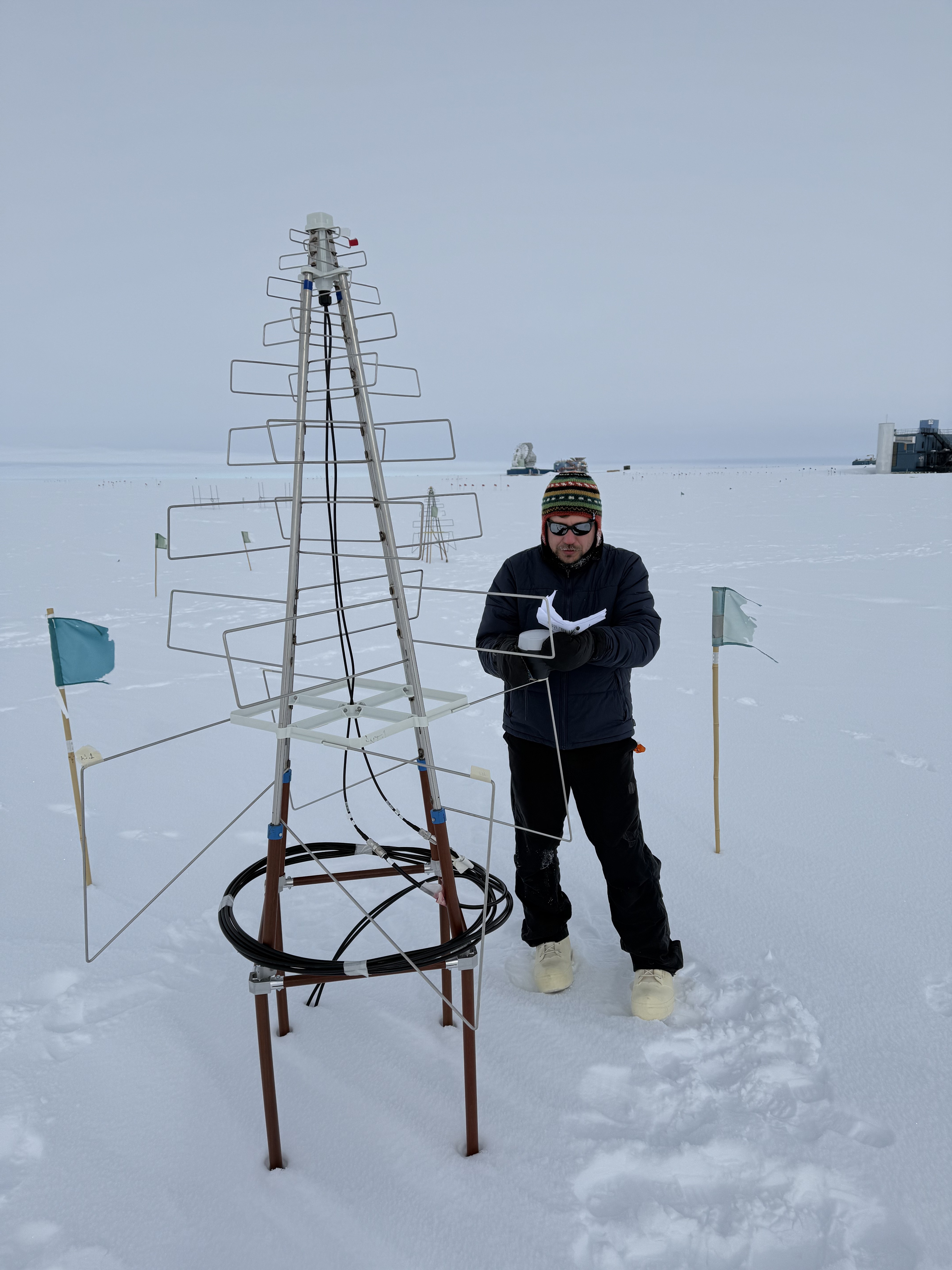}
        \label{fig:Dep2025}
    \end{subfigure}
     \vspace{-0.7cm}
    \caption{Left: The layout of the two newly deployed stations and existing Station 0 (formerly Prototype Station) of the Surface Enhancement with the upgraded antennas within the IceTop array. The IceCube Laboratory (ICL), the data hub for the Neutrino Observatory and IceTop is shown in dark blue.
    Right: A picture of A. Novikov conducting the GPS Survey on the antennas of Station 0 on December 18, 2024 before replacement.}
    \label{fig:Newlydeployedstations}
\end{figure}

\begin{figure}[h]
    \vspace{-0.5cm}
    \centering
    \begin{subfigure}{0.45\linewidth}
        \centering
        \includegraphics[width=\linewidth]{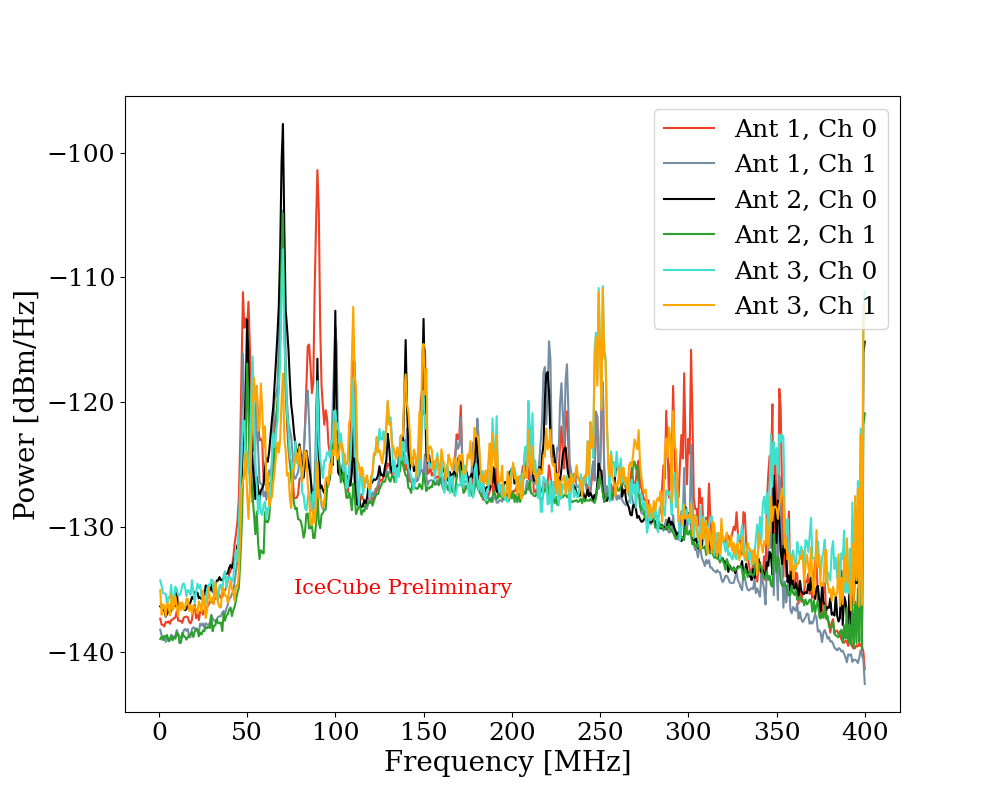}
        \label{fig:saps}
    \end{subfigure}
    \hfill
    \begin{subfigure}{0.45\linewidth}
        \centering
        \includegraphics[width=\linewidth]{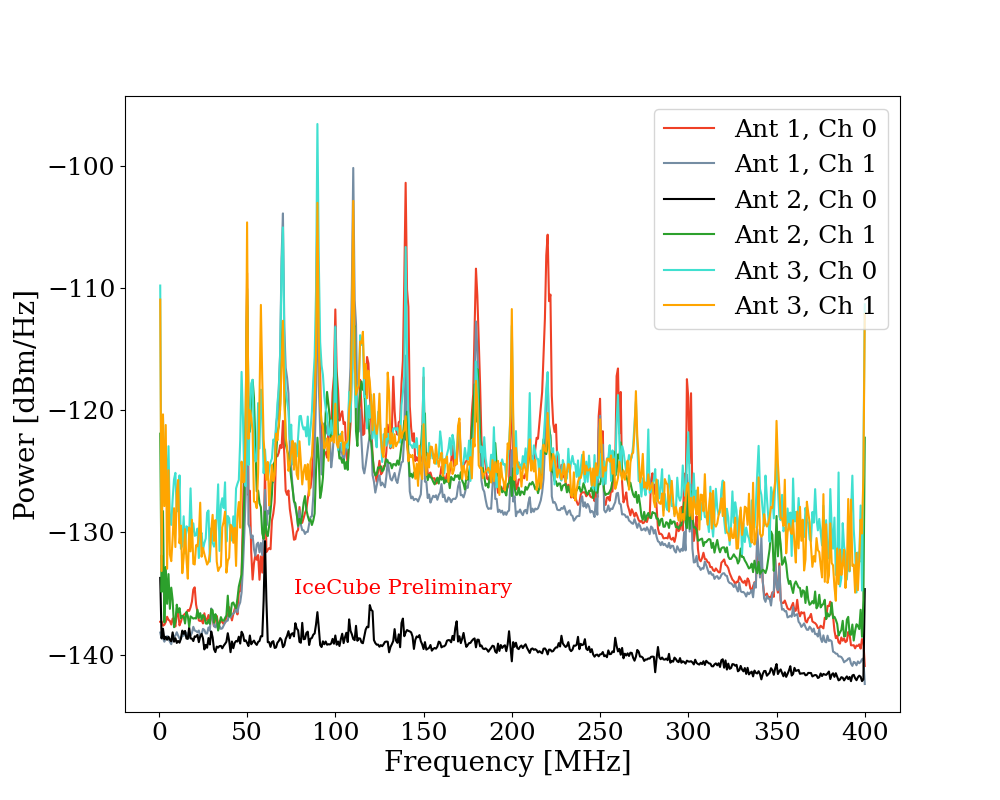}
        \label{fig:sa21}
    \end{subfigure}
    \hfill
        \begin{subfigure}{0.45\linewidth}
        \vspace{-0.725cm}
        \centering
        \includegraphics[width=\linewidth]{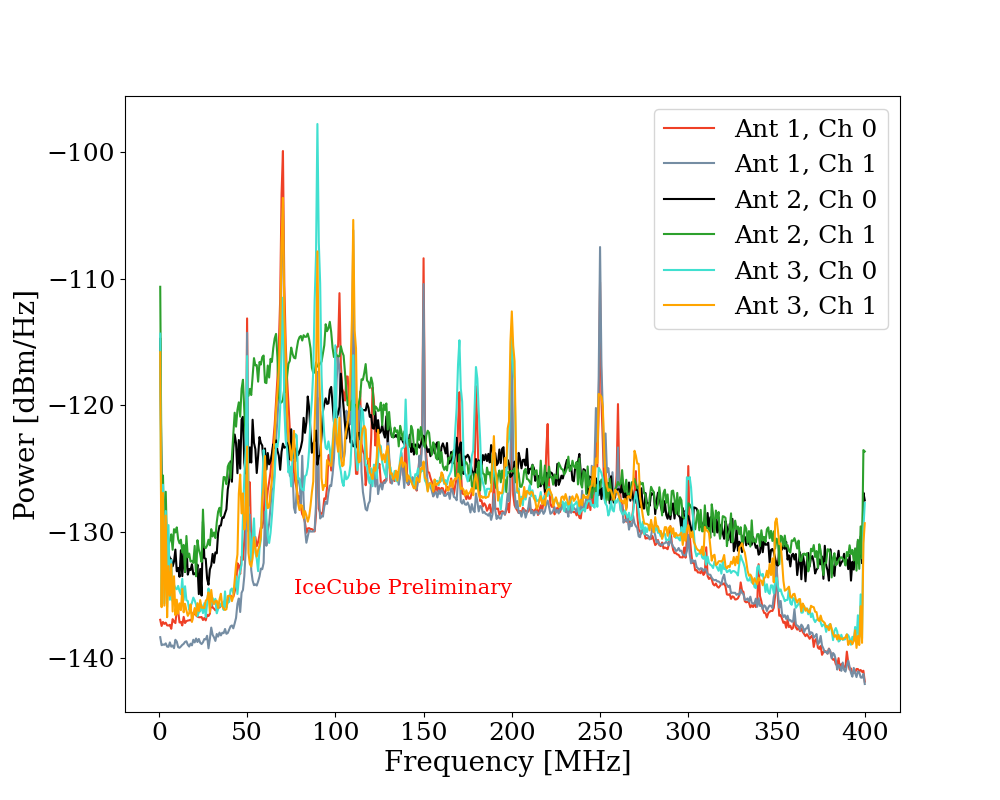}
        \label{fig:sa25}
    \end{subfigure}
    \vspace{-1cm}
    \caption{Average background spectra of three antennas for all the three stations deployed at the South Pole for one day of data (11th February, 2025). Left: Station 0 (formerly Prototype Station), Right: Station 21, Center below: Station 25.}
    \label{fig:FreqSpec3stations} 
\end{figure}

The field season 2024/25 was a major milestone for the surface enhancement with two newly deployed stations after several years of COVID related delays. All components of the stations, the scintillation detectors, radio antennas and the data acquisition systems were tested, calibrated and shipped to the South Pole. Trenching, cabling and the successful deployment  of two new complete stations were carried out as well as the upgrade of the existing antennas of Station 0 (formerly prototype station). During deployment, data from the antennas were read out and checked. To reduce heating related issues that affected the data from previous years, two main changes were made. The White Rabbit node, that was previously installed inside the hybrid data acquisition system, is now externally placed in the Fieldhub. Some part of the insulation foam from the Fieldhub of the Station 0 was also removed. Higher temperatures were known to cause readout errors with the scintillation detectors and affected the quality of radio data \cite{thesisRox,thesisNoah}. With the removal of the insulation, the temperatures inside the Fieldhub of Station 0 have reduced approximately by 15\degree\,C. The Fieldhubs for the newer stations are larger to accommodate the electronic junction boxes and the external White Rabbit Node.
The positions of the new stations on the IceTop array is shown on left in Fig.~\ref{fig:Newlydeployedstations} with a picture on the right of A. Novikov taking GPS survey measurements of Station 0 to confirm positions of the detectors before replacing antennas.
Nine antennas and their polarization channels with the exception of one channel seem to be working well. The power spectral density plots over averaged background data over an entire day is shown in Fig.~\ref{fig:FreqSpec3stations}. The spikes in the frequency spectra vary throughout the year but the measured spectrum follows the expected Galactic noise model. Integration of the data into the processing chain is ongoing.

\section{Results and Conclusion}\label{sec2}
The successful identification of 233 air showers in the 2023 dataset, as well as during a brief period in 2022 \cite{ARENAMeg}, demonstrates that the current radio reconstruction method is well-established and readily applicable to data from all years. The events have been verified and checked for coincidence with IceTop and checked for directional reconstruction. These events are ready to perform analysis on, including the estimation of $X_\mathrm{max}$. With the deployment of two further stations we can expect an increase in the number of measured air showers, further improvement and increase in all measurable quantities and improvements in resolution of the directional reconstruction.
\clearpage
\bibliographystyle{ICRC}
\bibliography{references}

%

\clearpage

\input{authorlist_IceCube.tex}

\end{document}

%% file: ICRCdetails.tex
\ConferenceLogo{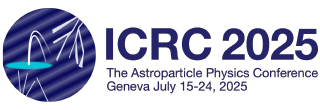}

\FullConference{39th International Cosmic Ray Conference (ICRC2025)\\
 15–24 July 2025\\
Geneva, Switzerland\\}

%% file: authorlist_IceCube.tex
\section*{Full Author List: IceCube Collaboration}

\scriptsize
\noindent
R. Abbasi$^{16}$,
M. Ackermann$^{63}$,
J. Adams$^{17}$,
S. K. Agarwalla$^{39,\: {\rm a}}$,
J. A. Aguilar$^{10}$,
M. Ahlers$^{21}$,
J.M. Alameddine$^{22}$,
S. Ali$^{35}$,
N. M. Amin$^{43}$,
K. Andeen$^{41}$
% ...
\\
\\
% ...
$^{10}$ Universit{\'e} Libre de Bruxelles, Science Faculty CP230, B-1050 Brussels, Belgium \\
% ...
$^{16}$ Department of Physics, Loyola University Chicago, Chicago, IL 60660, USA \\
$^{17}$ Dept. of Physics and Astronomy, University of Canterbury, Private Bag 4800, Christchurch, New Zealand \\
% ...
$^{21}$ Niels Bohr Institute, University of Copenhagen, DK-2100 Copenhagen, Denmark \\
$^{22}$ Dept. of Physics, TU Dortmund University, D-44221 Dortmund, Germany \\
% ...
$^{35}$ Dept. of Physics and Astronomy, University of Kansas, Lawrence, KS 66045, USA \\
% ...
$^{39}$ Dept. of Physics and Wisconsin IceCube Particle Astrophysics Center, University of Wisconsin{\textemdash}Madison, Madison, WI 53706, USA \\
% ...
$^{41}$ Department of Physics, Marquette University, Milwaukee, WI 53201, USA \\
% ...
$^{43}$ Bartol Research Institute and Dept. of Physics and Astronomy, University of Delaware, Newark, DE 19716, USA \\
% ...
$^{63}$ Deutsches Elektronen-Synchrotron DESY, Platanenallee 6, D-15738 Zeuthen, Germany \\
$^{\rm a}$ also at Institute of Physics, Sachivalaya Marg, Sainik School Post, Bhubaneswar 751005, India \\
% ...

\subsection*{Acknowledgments}

\noindent
The authors gratefully acknowledge the support from the following agencies and institutions:
USA {\textendash} U.S. National Science Foundation-Office of Polar Programs,
U.S. National Science Foundation-Physics Division,
U.S. National Science Foundation-EPSCoR,
U.S. National Science Foundation-Office of Advanced Cyberinfrastructure,
Wisconsin Alumni Research Foundation,
Center for High Throughput Computing (CHTC) at the University of Wisconsin{\textendash}Madison,
Open Science Grid (OSG),
Partnership to Advance Throughput Computing (PATh),
Advanced Cyberinfrastructure Coordination Ecosystem: Services {\&} Support (ACCESS),
Frontera and Ranch computing project at the Texas Advanced Computing Center,
U.S. Department of Energy-National Energy Research Scientific Computing Center,
Particle astrophysics research computing center at the University of Maryland,
Institute for Cyber-Enabled Research at Michigan State University,
Astroparticle physics computational facility at Marquette University,
NVIDIA Corporation,
and Google Cloud Platform;
Belgium {\textendash} Funds for Scientific Research (FRS-FNRS and FWO),
FWO Odysseus and Big Science programmes,
and Belgian Federal Science Policy Office (Belspo);
Germany {\textendash} Bundesministerium f{\"u}r Forschung, Technologie und Raumfahrt (BMFTR),
Deutsche Forschungsgemeinschaft (DFG),
Helmholtz Alliance for Astroparticle Physics (HAP),
Initiative and Networking Fund of the Helmholtz Association,
Deutsches Elektronen Synchrotron (DESY),
and High Performance Computing cluster of the RWTH Aachen;
Sweden {\textendash} Swedish Research Council,
Swedish Polar Research Secretariat,
Swedish National Infrastructure for Computing (SNIC),
and Knut and Alice Wallenberg Foundation;
European Union {\textendash} EGI Advanced Computing for research;
Australia {\textendash} Australian Research Council;
Canada {\textendash} Natural Sciences and Engineering Research Council of Canada,
Calcul Qu{\'e}bec, Compute Ontario, Canada Foundation for Innovation, WestGrid, and Digital Research Alliance of Canada;
Denmark {\textendash} Villum Fonden, Carlsberg Foundation, and European Commission;
New Zealand {\textendash} Marsden Fund;
Japan {\textendash} Japan Society for Promotion of Science (JSPS)
and Institute for Global Prominent Research (IGPR) of Chiba University;
Korea {\textendash} National Research Foundation of Korea (NRF);
Switzerland {\textendash} Swiss National Science Foundation (SNSF).